\documentclass[conference]{IEEEtran}
\ifCLASSINFOpdf
\else
\fi
\usepackage{amsmath}
\usepackage{graphicx}
\usepackage{amsfonts,amssymb}
\usepackage{mathrsfs}
\usepackage{tabularx}
\usepackage{enumerate}
\usepackage{booktabs}
\usepackage{algorithm}
\usepackage{algorithmic}
\usepackage{caption}
\usepackage{subfigure}
\usepackage{graphicx}
\usepackage{epstopdf}
\usepackage{cite}
\usepackage{changepage}
\usepackage{amsmath}
\newtheorem{theorem}{Theorem}
\newtheorem{lemma}{Lemma}

\begin{document}
\title{Cache-Enabled Dynamic Rate Allocation via Deep Self-Transfer Reinforcement Learning}


\author{\IEEEauthorblockN{Zhengming Zhang, Yaru Zheng, Meng Hua, Yongming Huang and Luxi Yang}
\IEEEauthorblockA{School of Information Science and Engineering, Southeast University, Nanjing 210096, China\\
Email: {{\{zmzhang, yrzheng, mhua, huangym and lxyang\}}@seu.edu.cn}}}

%\author{\IEEEauthorblockN{Zhengming Zhang}
%\IEEEauthorblockA{School of Electrical and\\Computer Engineering\\
%Georgia Institute of Technology\\
%Atlanta, Georgia 30332--0250\\
%Email: http://www.michaelshell.org/contact.html}
%\and
%\IEEEauthorblockN{Homer Simpson}
%\IEEEauthorblockA{Twentieth Century Fox\\
%\and
%\IEEEauthorblockN{James Kirk\\ and Montgomery Scott}
%\IEEEauthorblockA{Starfleet Academy\\
%San Francisco, California 96678--2391\\

\maketitle

\begin{abstract}
Caching and rate allocation are two promising approaches to support video streaming over wireless network. However, existing rate allocation designs do not fully exploit the advantages of the two approaches. This paper investigates the problem of cache-enabled QoE-driven video rate allocation problem. We establish a mathematical model for this problem, and point out that it is difficult to solve the problem with traditional dynamic programming. Then we propose a deep reinforcement learning approaches to solve it. First, we model the problem as a Markov decision problem. Then we present a deep Q-learning algorithm with a special knowledge transfer process to find out effective allocation policy. Finally, numerical results are given to demonstrate that the proposed solution can effectively maintain high-quality user experience of mobile user moving among small cells. We also investigate the impact of configuration of critical parameters on the performance of our algorithm.
\end{abstract}


\IEEEpeerreviewmaketitle

\section{Introduction}
Due to the exponential growth of the number of smart mobile equipments and innovative high-rate mobile data services (such as videos streaming for mobile gaming and road condition monitoring), the networks should accommodate the overwhelming wireless traffic demands \cite{Zheng201510, Jaber20165G}. In fact it can be predicted that the mobile video traffic will be over one third of mobile data traffic by the end of 2018 \cite{index2013global}. People's demand for video loading speed and video clarity is endless, thus the viewers' Quality-of-Experience (QoE) is an important indicator of the mobile communication network performance.

Mobile users often encounter video lag or sudden blur when using mobile devices to watch videos. The reason for these two cases is that the video is divided into small video chunks (usually of the same length) by special algorithms \cite{Carlsson2017Optimized}. If the current network is of low quality, the video sender will reduce the video resolution of the next few seconds to ensure that users can continue to watch the video \cite{Akhshabi2011An, Li2014Probe}. In this way, the quality can not be guaranteed. If the user fasts forward the video, and the paragraph has not been loaded, then the video playback will be interrupted until the corresponding paragraph is cached. The goal of using the adaptive bit rates allocation algorithm is to provide a better user experience and to reduce network bandwidth usage. We will focus on three factors that affect the user experience: quality (quantified by video rates), packet dropping and video frozenness duration.

Recently, caching at base stations has been used as a promising approach for video streaming \cite{7539325}. Caching during off-peak times can bring popular contents closer to users, and hence improves users' QoE \cite{7915753}. Although caching effectiveness has been extensively explored in the literature \cite{7539325}, \cite{7915753},\cite{6665021}, they do not consider the influence of bit rate allocation to the effectiveness of caching.

In this paper, we adapt the ideas underlying the success of QoE-driven dynamic video rate allocation to the cache-enabled wireless network. There are two important and practical assumption in our paper.

\textbf{Assumption 1 :} The bit rate received by the user is lied in a large discrete space which almost can be considered as a continuous range.

\textbf{Assumption 2 :} The central network has finite caching capability but knows the size and status of the buffer of the user's mobile device \cite{Chen2016QoE}, \cite{7996600}.

Most of the existing studies assume that the bit rate is in discrete space, and the dimension of the space is low so that all the values can be traversed within a finite time \cite{Li2014Probe,Chen2016QoE,7996600,Wang2017A}. This can be achieved by uniformly quantizing the candidate regions where the bit rate is located. The main reason for doing so is to simplify the complex real-world model and to facilitate the solution of the problem. In practice, however, the bit rate received by the user is lied in a large discrete space which almost can be considered as a continuous range.

In our work, we maximize the bit rate of video chunks that are actually consumed in a period of time, and minimize packet dropping and video frozenness duration. We track the network capacity and the buffer state like \cite{Chen2016QoE}. However, unlike previous studies, the capacity of the network and buffer state are considered as continuous values. We formulate the problem as a Markov Decision Process (MDP), then we present a deep Q-learning approach with normalized advantage functions and special knowledge transfer process to solve it.

The main contributions of this paper are summarized as follows:

(i) We define a more realistic QoE criteria that consider the total accumulation bit rates, packet dropping and video freeze duration for cache-enabled video streaming.

(ii) We convert the allocation problem into a Markov decision problem which extends the network capacity and candidate bit rate from discrete space to continuous space. And we point that the states transition probability cannot be obtained directly.

(iii) We propose a way to go beyond the traditional algorithms we call it deep self-transfer reinforcement learning. We get (sub)optimal dynamic video rate allocation policy by using this approach.

The rest of this paper is organized as follow. In Section II, we introduce existing rate adaptation approaches. Section III the system model is presented. Section IV, we proposed the deep reinforcement learning algorithms in which the solution of bitrate allocation policy for video streaming is verified. Finally, numerical results and discussion are presented in Section V, and a conclusion is reached in Section VI.

\section{Related Work}

Improving the QoE of adaptive video streaming has been the main focus of many researchers in recent years. Although caching methods used to optimize the video transmission, the audience's requirements for the video quality are endless, the sites providing wireless video services have to compromise between the quality and transmission speed.

\textbf{Model :} Numerous models have been proposed to address real-time adaptation of multimedia contents over wireless channels for a better QoE. A novel cache-enabled video rate control model that balances the needs for video rate smoothness and energy cost saving is proposed in \cite{7996600}. In this work, they consider the playback rate in their QoE model, but the transmission delay and caching delay are ignored. A logarithmic QoE model derived from experimental results is used in \cite{6463456} and they formulate the content cache management for video streaming into a convex optimization to give their problem an analytical framework. A QoE-aware video rate adaptation model is also proposed in \cite{Chen2016QoE}. In this work the user's buffer state is modeled as a vector and the value of network capacity is discretized as a finite set. To describe the dynamic evolution of the entire network, a large number of studies have modeled the problem as a Markov decision problem \cite{Chen2016QoE}, \cite{Bao2015Bitrate,Bokani2016Implementation,Pesce2014A}. We emphasize that the above studies ignore the impact of delay on QoE and cache performance and their MDP models are built in discrete action space and discrete state space.

\textbf{Algorithm :} Convex optimization algorithms are used in \cite{7996600, 6463456} to solve their allocation problems. Their algorithms are simple and efficient but are not valid for complex and dynamic scenes. Optimal policy for dynamic MDP model in \cite{Bao2015Bitrate} is based on a fast heuristic. The state transition function is given in detail (this is not practical in the real world) and standard value iteration method \cite{puterman2014markov} is used to obtain the optimal policy. In \cite{Pesce2014A} the optimal problem is solved via dynamic programming \cite{bertsekas1995dynamic}. In \cite{Chen2016QoE} similar approaches which based on value iteration and Q function are used to solve the MDP problem. It is worth emphasizing that the methods used in these studies are suitable to solve discrete dynamic programming problems and are not usbale for large scale decision space and continuous space.

Our study considers the cache-enabled video rate allocation MDP problem in continuous state space and action space, and we solve it using continuous deep Q-learning approach. This approach has three key technologies. They are continuous deep reinforcement learning \cite{Lillicrap2015Continuous, Schulman2015Trust, gu2016continuous}, imitation learning \cite{Schaal1999Is, Duan2017One} and transfer learning \cite{Rusu2016Progressive}.

\textbf{Continuous deep reinforcement learning:} Deep reinforcement learning has received considerable attention in recent years and has been applied to a wide variety of real-world robotic control tasks \cite{Lillicrap2015Continuous} including learning policies to play go game \cite{Silver2016Mastering}. To incorporate the benefits of value function estimation into continuous deep reinforcement learning, policy network and value network should be established \cite{Schulman2015Trust}. A different point of view is put forward in \cite{gu2016continuous}, the authors complete the task by learning a single network that outputs both the value function and policy. We are inspired by this work to create a deep Q-learning approach to solve our problem without any prior knowledge about the state transition.

\textbf{Imitation learning:} This kind of algorithms consider the problem of acquiring skills from observing demonstrations. Behavioral cloning and inverse reinforcement learning are two main lines of work within imitation learning. There two kinds of imitation learning can improve the learning efficiency of reinforcement learning \cite{gu2016continuous, Duan2017One}.

\textbf{Transfer learning:} This kind of algorithms consider the problem of learning policies with applicability and re-use beyond a single task. They can acquire a multitude of skills faster than what it would take to acquire each of the skills independently, and let the artificial intelligence revolution from virtual to reality \cite{Rusu2016Progressive}.

While there past works have led to a wide range of impressive results, they consider each skill separately, and do not fully exploit the advantages of the three key technologies. These led us to integrate these technologies with their respective advantages and get a more powerful algorithm.

\section{Model And Problem Formulation}

\begin{figure}
\centering
\resizebox{8.0cm}{3.5cm}{\includegraphics{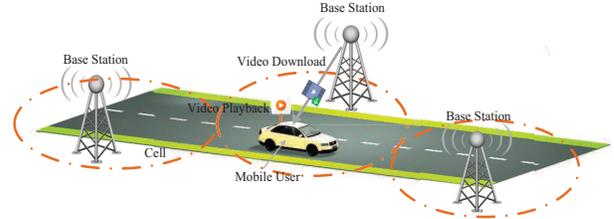}}
\caption{Video streaming architecture.}
\end{figure}
A highway high-mobility scenario with multiple base stations (BSs) deployed is shown in Fig. 1. We consider the downlink of a cache-enabled content-centric wireless network which provides full coverage of all driving sections and high data rate in each of the coverage areas. The BSs are equipped with cache storing some subsets of video contents. We assume the contents stored in the cache are given \cite{Liu2015Energy}. The video stream is divided into $N$ chunks and the $m$th chunk has a length of $\Delta T_m$ seconds. The network consisting of \(F\) overlaid small cell base stations and each of them is connected to the video server via optical fibers. The indices of \(F\) BSs are orderly included in a set \({\mathcal{B}} = \{ 1,2, \cdots ,F\}\). The caching system contains $L$ files ${{\cal W}} = \{ W_1 ,W_2 , \cdots ,W_L \}$. The user moving on a segment of highway which is covered by \({\cal B}\). Both BSs and mobile user are equipped with directional antennas. We define the following variables to describe our problem: state space, action space, state transition probabilities and reward function. The state space is a finite set of states. The action space is a finite set of actions. The state transition probabilities is the probability that action $a$ in state $s$ at time $t$ will lead to state $ s^{'}$. And the reward function is the immediate reward received after transitioning from state $s$ to state $s^{'}$, due to action $a$ .

\subsection{The state space}
Consider time slots indexed by \(T = \{ t_1 ,t_2 , \cdots ,t_\aleph  \}\) where \(\aleph\) is the number of slots. The time that the system stays at each slot is \(\Delta t_x  = t_{x + 1}  - t_x\). We assume that the network capacity is a random variable following a certain distribution and its value at time $t_x$ is $C^{t_x} \in \mathbb{R}^{+}$. Let $n^{t_x}$ as the requested content and $\textbf{\emph{B}}(t_x) \in \mathbb{R}^{M}$ as the buffer state where $M$ is large enough. The element $b_{i}(t_x)$ of $\textbf{\emph{B}}(t_x)$ is the bitrate of the $i$th video chunk and it must meet two conditions \cite{Chen2016QoE}, the first is FIFO constraint which means if $b_{i}=0$ then $b_{j}=0,j>i$, and the second is buffer length constraint which means \(
\sum\nolimits_i {b_i }<U\), where $U$ is the total size of the buffer.

Let \(\textbf{\emph{S}}^{t_x}\) denote the current state of the system at time $t_x$, the state \(\textbf{\emph{S}}^{t_x}\) is the combination of \(C^{t_x}\), $n^{t_x}$ and \(\textbf{\emph{B}}^{t_x}\). The state space $\mathcal{S}$ is composed of all the possible combination of the states.
\begin{equation}
{{\mathcal{S}}} = \left\{ {\textbf{\emph{S}}^{t_x}  = (C^{t_x} ,n^{t_x},\textbf{\emph{B}}^{t_x}) | {C^{t_x}  \in \mathbb{R}^{+},n^{t_x} \in {\cal W},\textbf{\emph{B}}^{t_x}  \in \mathbb{R}}^{M} }\right \}
\end{equation}

\subsection{The action space}

When the system is in state \(\textbf{\emph{S}}^{t_x} \), the set of possible actions $\textbf{\emph{A}}^{t_x}_j$ is the set of bit rates which would be allocated to the user from the base station $j$. Specifically, \(\textbf{\emph{A}}^{t_x}_j = (b_1 ,b_2 , \cdots ,b_k )\), where $b_{i}=b$ is the bit rate of the $i$th video chunk. The bit rate $b$ should satisfy \(b_{\min }  \le b \le b_{\max }\), and the number of video chunks that can be successfully transmitted is
\begin{equation}
k = \left\lfloor {\frac{{C^{t_x}}}{b}} \right\rfloor
\end{equation}
where \(\left\lfloor {\,} \right\rfloor\) indicates a round-down operation. Therefore, the action space can be defined as the possible combination of $\textbf{\emph{A}}^{t_x}_j$ i.e., \({{\cal A}} =  \cup \textbf{\emph{A}}^{t_x}_j\).

\subsection{The state transition probabilities}
The state transition probability from state \(S^{t_j }\) to state \(S^{t_k }\) can be given by
\begin{equation}
p_{j,k}  = p_{j,k} (C^{t_x} , n^{t_x}, \textbf{\emph{B}} ^{t_x} )
\end{equation}

\emph{Remark 1:} Although the expression of the transition probability is given here, the mapping relation cannot be used directly to solve problems, because it is not available in the real world and is difficult to estimate accurately.

\subsection{The reward function}

A reward function defines the goal in a reinforcement learning problem. In the rate allocation problem, the objective is to select proper action for each system state to optimal the overall performance of the network and video streaming quality. We divide the reward function into two parts, cache miss cost and video streaming quality. Cache miss cost measures the cost of downloading data from the server. Video streaming quality is used to measure the delay caused by video wireless transmission.

\textbf{Cache miss cost:} First, we consider the overhead caused by data download from the remote sever. Let $c(n^{t_x})$ denote the cost for fetching content $n^{t_x}$ via the backhaul link, depending on the content size. Then the cache miss cost is given by
\begin{equation}
C^{t_x } _{n,dl}  = {\textbf{1}}(n^{t_x} \notin {{\cal C}})c(n^{t_x})
\end{equation}
where $\cal C$ is the cache set, ${\textbf{1}}( \cdot )$ denotes the indicator function.

\textbf{Video streaming quality:} Now we consider the reward of the transmission. There three kinds of rewards of it, video playback quality $u^{t_x}$, packet loss cost \(C^{t_x } _{loss}\) and video freeze cost \(C^{t_x } _{freeze}\). The video playback quality is represented by the accumulated rate of the video chunks watched in the duration  \(\Delta t_x \) by the viewer.
\begin{equation}
u^{t_x } (S^{t_x } ,\textbf{\emph{A}}^{t_x } ) = \sum\limits_{i = 1}^{\widetilde k} {b_i }
\end{equation}
where ${\widetilde k}$ is number of video chunks pushed from the buffer in \(\Delta t_x \).
%Now we consider the cost of the quality of the video resources transmitted to the user, there are two kinds of costs, packet loss cost \(C^{t_x } _{loss}\) and video freeze cost \(C^{t_x } _{freeze}\).
For the user in the cell $C_j$ the base station \(B_j\) will transfer the video clip of size \(Z_{j}\) to the user, and \(Z_{j}\) can be given as $Z_{j}  = \sum\limits_{m = 1}^{M} {b_{m} {\Delta T_m} }$. Assume that the buffer size of the user at that time is \(U^{t_x}\leq U \), and $T_{t_x}$ is the time spent emptying $U^{t_x }$. Then the packet loss cost and the video freeze cost can be given by
\begin{equation}
C_{loss}^{t_x }  = \left\{ \begin{array}{l}
 \sum\limits_{m=\tau  + 1}^{ M } {\Delta T_m } ,\ if\ Z_{j}  + U^ {t_x } > U  \\
 0,\ otherwise \\
 \end{array} \right.
\end{equation}
\begin{equation}
\begin{split}
C_{freeze}^{t_x }  = \left\{ \begin{array}{l}
 \Delta t_x  -{\textbf{1}(n^{t_x} \notin {{\cal C}})\widetilde T_n}- T_{t_x} - \sum\limits_{m = 1}^\tau  {\Delta T_m } ,\\\quad~~ if\ \Delta t_x  > {\textbf{1}(n^{t_x} \notin {{\cal C}})\widetilde T_n}+T_{t_x}+ \sum\limits_{m = 1}^\tau  {\Delta T_m }  \\
 0,\ otherwise \\
 \end{array} \right.
\end{split}
\end{equation}
where \(\tau\) is the number of video blocks that can be received successfully by the user, $\widetilde T_n$ is the time taken to download the video $n^{t_x}$ from the remote server.

Finally we combine cache miss cost and video streaming quality as our QoE model to get the reward of the user at \(t_x\).
\begin{equation}
r^{t_x}_{\rm QoE} = \sum\limits_{i = 1}^4 {\lambda _i r_i }
\end{equation}
where $r_1  =  - C_{n,dl}^{t_x }$, $r_2  = u^{t_x }$, $r_3  =  - C_{loss}^{t_x } $, $r_4  =  - C_{freeze}^{t_x }$  and \(\lambda _i\) are associated weights for the $r_i$.  Given the cost of each slot, the total reward throughout the whole process is
\begin{equation}
R =  \sum\limits_{x = 1}^\aleph  {\sum\limits_{n = 1}^N {\gamma _j ^{t_x } r^{t_x } _{\rm QoE} } }
\end{equation}
%\begin{equation}
%R =  \sum\limits_{x = 1}^\aleph  { {\gamma  ^{t_x } r^{t_x } _{total} } }
%\end{equation}
where $N$ is the total number of requests sent by the user, \(\gamma _j ^{t_x }\) is the decay factor of the reward where $j \in {{\cal B}}$

\subsection{Problem formulation}
The global objective is to find the optimal bit rate allocation profile that maximizes the system reward throughout the whole process. The considered problem can be formulated as
\begin{equation}
\mathop {\min }\limits_{\textbf{\emph{A}}_j^{t_x } } \mathop {\lim \sup }\limits_{L \to \infty } \frac{-1}{{L\aleph}}\sum\limits_{x = 1}^\aleph {\sum\limits_{n = 1}^N {\gamma _j^{t_x } r_{{\rm{QoE}}}^{t_x } } _{\{ t_x  \in T,j \in {{\cal B}}\} } }
\end{equation}
where $L$ means that the system is running $L$ times and each time has $\aleph\in\mathbb{N}^+$ slots. Here $\textbf{\emph{A}}_j^{t_x }$ is the optimization variable, it is also action of the agent at time $t_x$. The system total reward is the objective function.

\textbf{Optimality analysis:} The cache-enabled bit rate allocation problem (10) is an infinite horizon cost MDP and it is well known to be difficult problem due to the curse of dimensionality. While dynamic programming approaches represent a systematic approach for MDPs, there are usually not practical due to the curse of dimensionality and they generally need the state transition probabilities which are not available in practice. Existing works do not solve problem (10) and the optimal allocation solution remains unknown and is highly nontrivial.

As for the problem (10), we can obtain the optimal bit rate allocation policy $\pi ^{\rm{*}}$ by solving the Bellman equation \cite{Bertsekas:1995:DPO:526593}.
\begin{lemma}[Bellman equation :]
There exist a value function $V( \cdot )$ and a scalar $\delta$ satisfying :
\begin{equation}
\delta  + V\left( \textbf{\emph{S}}^{t_x } \right)=\mathop {\max }\limits_{\textbf{\emph{A}} \in \cal A } \left\{ {\pi(\textbf{\emph{S}}^{t_x }|\textbf{\emph{A}}^{t_x })(r^{t_x}_{\rm QoE} +\gamma^{t_x}E[V(\textbf{\emph{S}}^{t_{x+1} })])} \right\}
\end{equation}
Where the expectation $E[\cdot]$ is with respect to the probability distribution of the request arrival $n^{t_x}$ and the network capacity $C^{t_x}$. $\delta  = V^{\rm{*}}$ is the optimal value to Problem and the optimal policy $\pi ^{\rm{*}}$ achieving the optimal value $V^{\rm{*}}$ is given by
\begin{equation}
\mu ^{\rm{*}}(\textbf{\emph{S}}^{t_x })=\arg \mathop {\max }\limits_{\textbf{\emph{A}} \in \cal A } \left\{ {\pi(\textbf{\emph{S}}^{t_x }|\textbf{\emph{A}}^{t_x })(r^{t_x}_{\rm QoE} +\gamma^{t_x}E[V(\textbf{\emph{S}}^{t_{x+1} })])} \right\}
\end{equation}
\end{lemma}
\emph{Proof :} Please refer to Appendix A.
%By Propositions 4.2.1, 4.2.3, and 4.2.5 in \cite{Bertsekas:1995:DPO:526593}, the optimal system reward of problem (10) is the same for all initial states and the solution $(\delta, V\left( \textbf{\emph{S}}^{t_x } \right))$ to the                                                                                            following Bellman equation exists.
%\begin{equation}
%\begin{split}
%& \delta  + V\left( \textbf{\emph{S}} \right)=
%\\&\mathop {\max }\limits_{\textbf{\emph{A}} \in \cal A } \left\{ {\pi(\textbf{\emph{S}}|\textbf{\emph{A}})(r^{t_x}_{\rm QoE} +\gamma^{t_x}\sum_{\textbf{\emph{S}}^{' } \in \cal {S} }p(\textbf{\emph{S}}^{' }|\textbf{\emph{S}},\textbf{\emph{A}}^)V(\textbf{\emph{S}}^{' }))} \right\}
%\end{split}
%\end{equation}
%The transition probability is given by
%\begin{equation}
%\begin{split}
%&p(\textbf{\emph{S}}^{' }|\textbf{\emph{S}},\textbf{\emph{A}}^)\buildrel \Delta \over =p(\textbf{\emph{S}}^{t_{x+1} }=\textbf{\emph{S}}^{' }|\textbf{\emph{S}}^{t_x }=\textbf{\emph{S}},\textbf{\emph{A}}^{t_x }=\textbf{\emph{A}})\\
%&=E_{n,C}[p(\textbf{\emph{S}}^{t_{x+1} }=\textbf{\emph{S}}^{' }|\textbf{\emph{S}}^{t_x }=\textbf{\emph{S}},\textbf{\emph{A}}^{t_x }=\textbf{\emph{A}}),n^{t_x}=n,C^{t_x}=C]
%\end{split}
%\end{equation}
%Substituting (14) into (13) leads to (11). We complete the proof.

\textbf{Algorithm design:} Closed solution almost does not exist to the infinite horizon cost MDPs without any knowledge of state transition probabilities, and standard brute-force algorithms are usually impractical for implementation due to the curse of dimensionality. Therefore, it is of great interest to develop low-complexity suboptimal solutions, which can adapt to different initial state and complex environment. A deep neural network is built to approximate the allocation policy whose input is any possible system state and output is appropriate action. It has three advantages, first it can use historical data to training off-line, second the trained deep neural network can be used online, third it has good generalization performance.

%\begin{equation}
%\mathop {\max }\limits_{\textbf{\emph{A}}^{t_x } _j } \sum\limits_{x = 1}^{\aleph} {\sum\limits_{n = 1}^N \gamma ^{t_x } } r^{t_x}_{total}{}_{\{ j \in {\cal B}\} }
%\end{equation}

\section{Solve Decision Process Problem}
In this section, we first introduce deep Q-learning that can work in continuous space. Then we embed a special imitation learning algorithm and a transfer learning algorithm into it to make our deep Q-learning more efficient.

\textbf{Continuous deep Q-learning:} The Q function $Q^\pi(\textbf{\emph{s}}_t,\textbf{\emph{a}}_t)$ is used in many reinforcement learning algorithms, it corresponds to a policy $\pi$ and is defined as the expected return after taking an action $\textbf{\emph{a}}_t$ in state $\textbf{\emph{s}}_t$ and following the policy $\pi$ thereafter.
\begin{equation}
{Q^\pi(\textbf{\emph{s}}_t,\textbf{\emph{a}}_t)} = \mathbb{E}_{r_{i \ge t} ,{\textbf{\emph{s}}}_{i>t}  \sim E,{\textbf{\emph{a}}}_{i > t} \sim \pi } \left[ {R_t | \textbf{\emph{s}}_t,\textbf{\emph{a}}_t  } \right]
\end{equation}
Q-learning uses the greedy policy \(\mu \left( {{\bf{s}}_t } \right) = \arg \max _{\textbf{\emph{a}}} Q\left( {{\textbf{\emph{s}}}_t,{\textbf{\emph{a}}}_t } \right)\). Let $\theta ^Q$ parametrize the Q function and the loss function is :
\begin{equation}
\begin{array}{l}
 L\left( {\theta ^Q } \right) = E_{{\textbf{\emph{s}}}_t  \sim \rho ^\beta  ,{\textbf{\emph{a}}}_t  \sim \beta ,r_t  \sim E} \left[ {\left( {Q\left( {{\textbf{\emph{s}}}_t ,{\textbf{\emph{a}}}_t |\theta ^Q } \right) - y_t } \right)^2 } \right]
 \end{array}
\end{equation}
where
\begin{equation}
y_t  = r^{t}_{\rm QoE}\left( {{\textbf{\emph{s}}}_t ,{\textbf{\emph{a}}}_t } \right) + \gamma Q\left( {{\textbf{\emph{s}}}_{t + 1} ,\mu \left( {{\textbf{\emph{s}}}_{t + 1} } \right)} \right)
\end{equation}
and $\rho$ is a stochastic distribution, $\beta$ is a different stochastic policy. We optimize the Q function by using gradient descent with $\frac{{\partial L(\theta ^Q )}}{{\partial \theta ^Q }}$:
\begin{equation}
\begin{array}{l}
 \frac{{\partial L(\theta ^Q )}}{{\partial \theta ^Q }} \\
  = E_{{\textbf{\emph{s}}}_t  \sim \rho ^\beta  ,{\textbf{\emph{a}}}_t  \sim \beta ,r_t  \sim E}\left[ {\nabla _{\theta ^Q } \left( {Q({{\textbf{\emph{s}}}}_t ,{{\textbf{\emph{s}}}}_t |\theta ^Q ) - y_t } \right)^2 } \right] \\
  = E_{{\textbf{\emph{s}}}_t  \sim \rho ^\beta  ,{\textbf{\emph{a}}}_t  \sim \beta ,r_t  \sim E}\left[ {\nabla _{\theta ^Q } Q({{\textbf{\emph{s}}}}_t ,{{\textbf{\emph{a}}}}_t |\theta ^Q )\left( {Q({{\textbf{\emph{s}}}}_t ,{{\textbf{\emph{a}}}}_t |\theta ^Q ) - y_t } \right)} \right] \\
 \end{array}
\end{equation}

To use Q-learning for continuous problem, we should also define the value function \(V^\pi  \left( {\textbf{\emph{s}}_t  } \right)\) and advantage function \(A^\pi  \left( {\textbf{\emph{s}}_t ,\textbf{\emph{a}}_t } \right)\):
\begin{equation}
\begin{array}{l}
V^\pi  \left( {{\textbf{\emph{s}}}_t } \right) = \int\limits_{\textbf{\emph{a}}} {\pi \left( {{\textbf{\emph{s}}}_t ,{\textbf{\emph{a}}}_t } \right)Q^\pi  \left( {{\textbf{\emph{s}}}_t ,{\textbf{\emph{a}}}_t } \right)d{\textbf{\emph{a}}}} \\
A^\pi  \left( {{\textbf{\emph{s}}}_t ,{\textbf{\emph{a}}}_t } \right) = Q^\pi  \left( {{\textbf{\emph{s}}}_t ,{\textbf{\emph{a}}}_t } \right)-V^\pi  \left( {{\textbf{\emph{s}}}_t } \right)\\
 \end{array}
\end{equation}
Then a neural network that separately outputs a value function term $V$ and an advantage term $A$ is used to represent the Q function, and the network is parameterized as a quadratic function [20]:
\begin{equation}
\begin{array}{l}
 Q\left( {{\textbf{\emph{s}}},{\textbf{\emph{a}}}|\theta ^Q } \right) = A\left( {{\textbf{\emph{s}}},{\textbf{\emph{a}}}|\theta ^A } \right) + V\left( {{\textbf{\emph{s}}}|\theta ^V } \right) \\
 A\left( {{\textbf{\emph{s}}},{\textbf{\emph{a}}}|\theta ^A } \right) =  - \frac{1}{2}\left( {{\textbf{\emph{a}}} - \mu \left( {{\textbf{\emph{s}}}|\theta ^\mu  } \right)} \right)^T {\textbf{\emph{P}}}\left( {{\textbf{\emph{s}}}|\theta ^P } \right)\left( {{\textbf{\emph{a}}} - \mu \left( {{\textbf{\emph{s}}}|\theta ^\mu  } \right)} \right) \\
 \end{array}
\end{equation}
where $\textbf{\emph{P}}(\textbf{\emph{s}}|\theta ^P)$ is a state dependent positive definite square matrix, and it is parameterized by $\textbf{\emph{P}}(\textbf{\emph{s}}|\theta ^P)=\textbf{\emph{L}}(\textbf{\emph{s}}|\theta ^L)^{T} \textbf{\emph{L}}(\textbf{\emph{s}}|\theta ^L)$, and $\textbf{\emph{L}}(\textbf{\emph{s}}|\theta ^L)$ is a lower triangular matrix. We have three variables that are parameterized differently ${\mu \left( {{\bf{s}}|\theta ^\mu  } \right)}$, ${V \left( {{\bf{s}}|\theta ^V  } \right)}$ and $\textbf{\emph{P}}(\textbf{\emph{s}}|\theta ^P)$, they are three neural network and we use target networks to combine them \cite{Mnih2015Human}.

\textbf{Imitation learning:} As large amounts of on-policy experience are required in addition to good off-policy samples, we should improve the sampling efficiency of the algorithm. For the video rate allocation problem in continuous action space some actions, for example, discrete values obtained by sampling in the action space at equal intervals \cite{Chen2016QoE}, are useful but not easily sampled. For this reason we propose an imitation learning we called it imitated sampling learning (ISL).
\begin{algorithm}
\caption{ISL}
\label{alg1}
\begin{algorithmic}[1]
 \STATE  \textbf{Initialize:} Initialize the set $I = \{ b_{\min } ,b_{\min }  + \Delta b,b_{\min }  + 2\Delta b, \cdots ,b_{\max } \}$ where $\Delta b = \frac{{b_{\max }  - b_{\min } }}{d}$ and $d$ is a fixed constant.\\
\STATE \textbf{Step 1:} Sample a random minbatch of \({\rm K}\) samples \(\{ (\textbf{\emph{S}}_{i} ,\textbf{\emph{A}}_{i} ,r_i ,\textbf{\emph{S}}_{i+1} )\} _{i = 1}^{\rm K}\) from replay buffer \(\Sigma\).\\
\STATE \textbf{Step 2:} Using there samples get a policy $\pi_\theta(\textbf{\emph{A}}_{i}|\textbf{\emph{S}}_{i})$\\
\STATE \textbf{Step 3:} Using the policy $\pi_\theta(\textbf{\emph{A}}_{i}|\textbf{\emph{S}}_{i})$ get new set $(\widetilde{\textbf{\emph{S}}}_{i},\textbf{\emph{A}}^{'}_{i})$ and discrete $\textbf{\emph{A}}^{'}_{i}$ i.e., \(
\widetilde{\textbf{\emph{A}}}_{i}  = (b, \cdots ,b),b \in I.\) Then get a new dataset $\widetilde{\Sigma}=(\widetilde{\textbf{\emph{S}}}_{i},\widetilde{\textbf{\emph{A}}}_{i},\widetilde r_i,\widetilde{\textbf{\emph{S}}}_{i+1})$ \\
\STATE \textbf{Step 4:} Return aggregate dataset $\Sigma\leftarrow\Sigma\cup\widetilde{\Sigma}$.
\end{algorithmic}
\end{algorithm}

In Algorithm 1, the \textbf{Initialize} gets a candidate bit rate set like \cite{Wang2017A}. \textbf{Step 2} provides a learned model $\pi_\theta$ and this model will be used get new samples. \textbf{Step 3} and \textbf{Step 4} can be seen as a sample imagination rollout \cite{gu2016continuous}. Q-learning inherently requires noisy on-policy actions to succeed. It implies that the policy must be allowed to make $¡°$its own mistakes$¡±$ during training, which might involve taking undesirable. ISL can avoid this problem while still allowing for a large amount of on-policy exploration is to generate synthetic on-policy trajectories under the learned model $\pi_\theta$.

ISL synthesizes a series of operations by discretizing $\textbf{\emph{A}}^t_j$ and use them to get a lot of state action trajectories and add them to the replay buffer. This will effectively augments the amount of experience available for Q-learning, and improve its data efficiency substantially.

\textbf{Self-Transfer learning:} Using pseudo-rewards in reinforcement learning can make other auxiliary predictions that serve to focus the agent on important aspects of the task. It also accelerates the acquisition of a useful representation \cite{Jaderberg2016Reinforcement}. Our QoE model will focus on different aspects when we choose different $\lambda_i$. This inspired us to choose different $\lambda_i$ to improve the generalization ability of our agent. And we give this ability to the original agent by using self-transfer learning (STL).

\begin{algorithm}
\caption{STL}
\label{alg1}
\begin{algorithmic}[1]
 \STATE  \textbf{Initialize:} Randomly initialize normalized networks $\widetilde{\textbf{\emph{P}}}({\textbf{\emph{s}}},{\textbf{\emph{a}}}|\theta ^{\widetilde{P}} )$, $\widetilde{V}({\textbf{\emph{s}}}|\theta ^{\widetilde{V}} )$, $\widetilde{\mu} ({\textbf{\emph{s}}}|\theta ^{\widetilde{\mu}}  )$ and a prediction network $G(\textbf{\emph{r}}|\theta^G)$.\\
\STATE \textbf{Step 1:} Copy $\widetilde{\textbf{\emph{P}}}({\textbf{\emph{s}}},{\textbf{\emph{a}}}|\theta ^{\widetilde{P}} )$ and $\widetilde{V}({\textbf{\emph{s}}}|\theta ^{\widetilde{V}} )$ to ${\textbf{\emph{P}}}({\textbf{\emph{s}}},{\textbf{\emph{a}}}|\theta ^P )$ and $V({\textbf{\emph{s}}}|\theta ^V )$. \\
\STATE \textbf{Step 2:} Using there $\widetilde{\mu} ({\textbf{\emph{s}}}|\theta ^\mu  )$ get dataset $\Sigma ^{'}  = \{ ({\textbf{\emph{S}}}_i ,{\textbf{\emph{A}}}_i ,{\textbf{\emph{r}}}_i  = (r_1 ,r_2 ,...,r_d ),{\textbf{\emph{S}}}_{i + 1} )\}$.\\
\STATE \textbf{Step 3:} Split each $\textbf{\emph{r}}_i$: $Split(r_i)=(e_ir_i,(1-e_i)r_i)$ where $e_i$ obeys a probability distribution $\beta$ and $\sum\limits_{i = 1}^d {e_i }=1$.\\
\STATE \textbf{Step 4:} Use $G(\textbf{\emph{r}}|\theta^G)$ for prediction error: $error_i  = (1 - e_i )r_i  - G(e_i r|\theta ^G )$. And train $G(\textbf{\emph{r}}|\theta^G)$ using $Split(r_i)$.\\
\STATE \textbf{Step 5:} Update $\widehat{r}_i=e_ir_i+((1-e_i)r_i-error_i)/2$. And get $\Sigma  = \{ ({\textbf{\emph{S}}}_i ,{\textbf{\emph{A}}}_i ,\sum\limits_{i = 1}^d {\widehat{r}_i},{\textbf{\emph{S}}}_{i + 1} )\}$.\\
\STATE \textbf{Step 6:} Use $\Sigma$ train $\widetilde{\textbf{\emph{P}}}({\textbf{\emph{s}}},{\textbf{\emph{a}}}|\theta ^A )$, $\widetilde{V}({\textbf{\emph{s}}}|\theta ^V )$ and $\widetilde{\mu}({\textbf{\emph{s}}}|\theta ^{\mu} )$. And update $\theta ^P \leftarrow \theta ^{\widetilde{P}}$ and $\theta ^V \leftarrow \theta ^{\widetilde{V}}$.
\STATE Return ${\textbf{\emph{P}}}({\textbf{\emph{s}}},{\textbf{\emph{a}}}|\theta ^P )$, $V({\textbf{\emph{s}}}|\theta ^V )$ and $\widetilde{\mu} ({\textbf{\emph{s}}}|\theta ^\mu  )$.
\end{algorithmic}
\end{algorithm}
In Algorithm 2, we construct different reward set by using different $e_i$ from a noise process $\mathfrak{N}$. For each reward $r_i$, $e_ir_i$ can be seen as the main part of the $\textbf{\emph{r}}_i$, and $(1-e_i)r_i$ is considered to be a noise added to $\textbf{\emph{r}}_i$, and vice versa. This inspired us to propose a method which called reward lifting scheme similar to lifting-schemed wavelet transform \cite{826776} to obtain the main components of the rewards. The reward lifting scheme has three main steps: split, predict, and update, which correspond to \textbf{Step 3}, \textbf{Step 4}, and \textbf{Step 5} in Algorithm 2, respectively.

STL synthesizes a series of pseudo-rewards and use them to train a virtual agent. Then copies its knowledge to the original agent. This can improve training efficiency and generalization performance. The training method to our deep self-transfer reinforcement learning is described in Algorithm 3.

\begin{algorithm}
\caption{Deep Self-Transfer Reinforcement Learning for Cache-Enabled Video Rate Allocation}
\label{alg1}
\begin{algorithmic}[1]
 \STATE  \textbf{Initialize:} Use STL and (18) initialize normalized Q network $Q\left( {{\textbf{\emph{s}}},{\textbf{\emph{a}}}|\theta ^Q } \right)$. And freeze ${\textbf{\emph{P}}}({\textbf{\emph{s}}},{\textbf{\emph{a}}}|\theta ^P )$ and $V({\textbf{\emph{s}}}|\theta ^V )$. Initialize target network $Q'$ with weight $\theta^{Q'}=\theta^{Q}$. Initialize replay buffer $\Sigma  = \emptyset$. \\
\STATE \textbf{For episode = 1 ,$\cdots$, \(\Gamma\) do:}
\STATE  \quad Initialize a random process \({{\cal N}}\) for action exploration.\\
\STATE \quad Get a initial observation state \(s^{t_1 }\).  \\
\STATE \quad \textbf{For \emph{x} = 1 ,$\cdots$, \(\aleph\) do:}
\STATE \qquad Select action  \(\textbf{\emph{A}}^{t_x}\) according to the current policy and exploration method $\textbf{\emph{A}}^{t_x}=\mu \left( {S^{t_x } |\theta ^\mu  } \right) + {{\cal N}}$. \\
\STATE \qquad Execute $\textbf{\emph{A}}^{t_x}$ and observe $r_{t_x}$ and $\textbf{\emph{S}}^{t_{x+1}}$.\\
\STATE \qquad Store $(\textbf{\emph{S}}^{t_x},\textbf{\emph{A}}^{t_x},r_{t_x},\textbf{\emph{S}}^{t_{x+1}})$ in $\Sigma$. \\
\STATE \qquad \textbf{If mod(episode $\cdot \aleph\ + \emph{x}$ ) = 0}
\STATE \qquad \quad Run Algorithm ISL.
\STATE \qquad \textbf{End If}
\STATE \qquad Sample a random minbatch of \({\rm K}\) samples \(\{ (\textbf{\emph{S}}_{i} ,\textbf{\emph{A}}_{i} ,r_i ,\textbf{\emph{S}}_{i+1} )\} _{i = 1}^{\rm K}\) from \(\Sigma\).\\
\STATE \qquad Set $y_i  = r_i  + \gamma V\left( {\textbf{\emph{S}}_{i + 1} |\theta ^{Q'} } \right)$. \\
\STATE \qquad Update $\theta ^{Q}$ by minimizing the loss function $
L = \frac{1}{{\rm K}}\sum\limits_{i = 1}^{\rm K} {\left( {y_i  - Q\left( {\textbf{\emph{S}}_{i} ,\textbf{\emph{A}}_{i} |\theta ^Q } \right)} \right)} ^2
$
\STATE \qquad Update the target network $ \theta ^{Q'}  = r\theta ^Q  + (1 - r)\theta ^{Q'} $. \\
\STATE \quad \textbf{End For}
\STATE \textbf{End For}
\end{algorithmic}
\end{algorithm}

In Algorithm 3, \(\Gamma\) is the total number of training times, \(r\) is the learning rate. We use Adam \cite{Kingma2014Adam} and the rectified non-linearity \cite{Glorot2010Deep} to learn the parameters for our neural network with a learning rate of $10^{-3}$ . We set the mini-batch sizes is 64 and the replay buffer size is $10^6$. Fig. 2 shows the workflow of the deep self-transfer reinforcement learning in detail. STL is first executed to get ${\textbf{\emph{P}}}({\textbf{\emph{s}}},{\textbf{\emph{a}}}|\theta ^P )$, $V({\textbf{\emph{s}}}|\theta ^V )$ and $\widetilde{\mu} ({\textbf{\emph{s}}}|\theta ^\mu  )$ with good generalization performance. This corresponds to the left part of Fig. 2. When the STL is sufficiently trained, the parameters of the STL will be passed to the deep neural network which located in the right of Fig. 2 and the ISL will run at the appropriate time. Fig. 3 is the structure of the deep self-transfer reinforcement learning neural network. ${\mu \left( {{\textbf{\emph{s}}}|\theta ^\mu  } \right)}$, ${V \left( {{\textbf{\emph{s}}}|\theta ^V  } \right)}$ and $\textbf{\emph{L}}(\textbf{\emph{s}}|\theta ^L)$, each of the three neural networks has three hidden layers with 128, 64 and 32 units, respectively. The output layer of ${\mu(\textbf{\emph{s}}|\theta ^\mu) }$ uses Sigmoid activation function. The training clock controls the time of knowledge transfer.
\begin{figure*}
\centering
\resizebox{12.0cm}{7.3cm}{\includegraphics{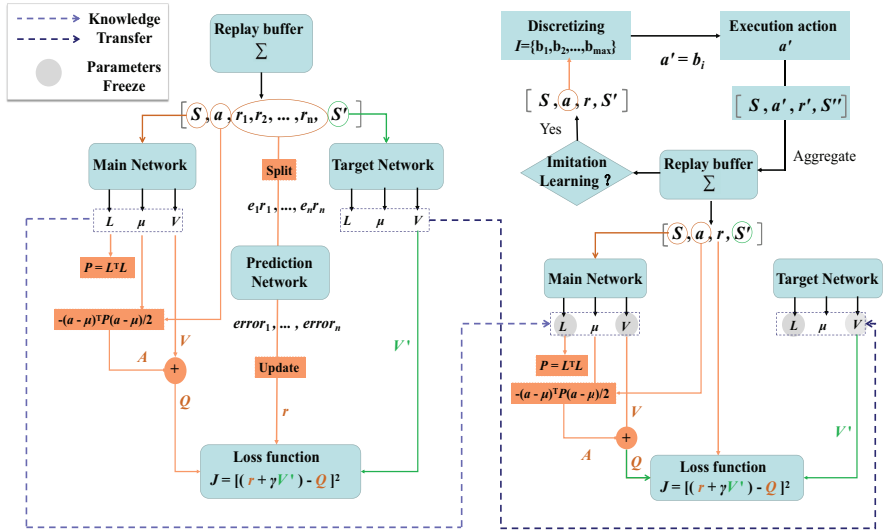}}
\caption{Deep self-transfer reinforcement learning.}
\end{figure*}

\begin{figure}
\centering
\resizebox{8.0cm}{6.5cm}{\includegraphics{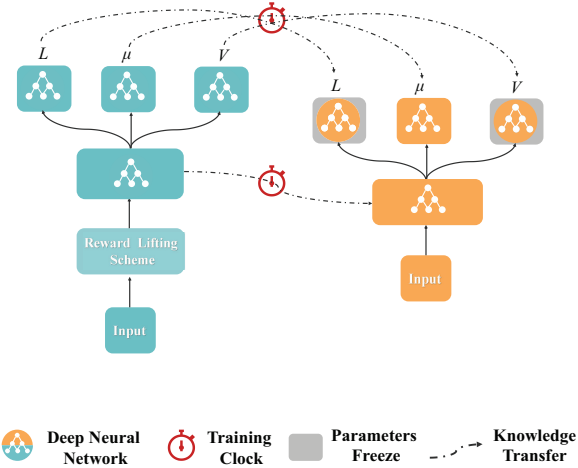}}
\caption{The structure of deep self-transfer reinforcement learning neural network.}
\end{figure}

%\begin{figure*}
%  \begin{minipage}[t]{0.5\linewidth}
%    \centering
%    \includegraphics[width=4.2in,height=2.4in]{Alg3.eps}
%    \caption{Deep Self-Transfer Reinforcement Learning.}
%  \end{minipage}%
%  \begin{minipage}[t]{0.5\linewidth}
%    \centering
%    \includegraphics[width=2.2in,height=1.8in]{Alg3Stucture.eps}
%    \caption{The Structure of Deep Self-Transfer Reinforcement Learning Neural Network}
%  \end{minipage}%
%\end{figure*}

\textbf{Analysis of Algorithm 3:} The training of the parameters of the deep neural network in algorithm 3 uses Adam based on the first-order gradient information which makes the neural network update sufficiently stable. Similar to \cite{Silver2016Mastering}, despite lacking any theoretical convergence guarantees, our method is able to train large neural networks using reinforcement learning signal and gradient descent in stable manner. There are two neural networks to calculate Q, $Q^{'\theta}$ and $Q^{\theta}$, only the target Q-network calculates Q values according to the next state with the target network's output. The using of target Q-network can cut off the dependencies of training samples and improve stability and convergence.

A major challenge of transfer learning in different tasks is catastrophic forgetting \cite{Mccloskey1989Catastrophic}. It means that the tendency of a neural network to completely and abruptly forget previously learned information upon learning new information. We teach a learned model parameter to a new task via lateral connection \cite{Rusu2016Progressive}. Since the trained model will be freezed in network design (line 1 in Algorithm 3), the way parameters are optimized for back-propagation does not affect the network that has been learned. This kind of neural network design naturally avoids the appearance of catastrophic forgetting.

The challenge of exploration bring us to learn a effective policy in continuous action spaces can be treated by adding noise sampled from a noise process ${\cal N}$ (line 6 in Algorithm 3). The main reason is the off-policies algorithms such as Algorithm 3 can treat the problem of exploration independently from the learning algorithm.

Notice that in order to solve (10) by using Bellman equation the knowledge of the transition probability is needed, but the formulated Markovian domain herein lacks the state transition mapping as state in Remark 1. Therefore, the traditional approaches cannot be used to solve our problem. Although we do not know the transition probability, our algorithm is still valid, we prove this by Theorem 1.
\begin{theorem}[Effectiveness]
The changes in the distribution of $n^{t_x}$ and $C^{t_x}$ do not cause Algorithm 3 to fail to update its parameters, no matter what the distribution function of $n^{t_x}$ and $C^{t_x}$ is Algorithm 3 is valid.
\end{theorem}

\emph{Proof :} Please refer to Appendix B.
%Since $n^{t_x}$ and $C^{t_x}$ affect the state transition probabilities, we let $d^{\mu}(\textbf{\emph{S}})=\sum_{x=0}^{\infty}\gamma^{t_x}p(\textbf{\emph{S}}^{t_x}=\textbf{\emph{S}}|\textbf{\emph{S}}^{t_0},\textbf{\emph{A}})$. Then use policy gradient \cite{Sutton1999Policy} and cost function $L$ (line 14 in Algorithm 3) we get
%\begin{equation}
%\frac{{\partial L}}{{\partial \theta }} = \sum\limits_{\textbf{\emph{S}}} {d^\mu  (\textbf{\emph{S}})} \sum\limits_{\textbf{\emph{A}}} {\frac{{\partial \mu (\textbf{\emph{S}},\textbf{\emph{A}})}}{{\partial \theta }}[A^{\mu}(\textbf{\emph{S}},\textbf{\emph{A}}) + V^{\mu}(\textbf{\emph{S}})]}
%\end{equation}
%The gradient is no terms of the form $\frac{{\partial d^\mu  (\textbf{\emph{S}})}}{{\partial \theta }}$, so the effect of policy changes on the distribution of states does not appear. We complete the proof.

We can find that the optimal policy $\pi ^*$ is existed from lemma 1. A popular measure of the performance of a reinforcement learning algorithm is its regret relative to executing the optimal policy in current MDP. The regret interests in the difference (in rewards during learning) between an optimal policy and a reinforcement learner:
\begin{equation}
\begin{array}{l}
 \Delta (\pi ,\pi ^* ) = \mathop {\lim }\limits_{T \to \infty } \frac{1}{T}\left( {\sum\limits_{x = 0}^T {\gamma _j^{t_x } r_{QoE}^{t_x } ({\textbf{\emph{S}}}^{t_x } ,{\textbf{\emph{A}}}^{t_x } |\pi )} -} \right. \\
 \quad \quad \quad \quad \quad \quad \quad \quad \left. { \sum\limits_{x = 0}^T {\gamma _j^{t_x } r_{QoE}^{t_x } ({\textbf{\emph{S}}}^{t_x } ,{\textbf{\emph{A}}}^{t_x } |\pi ^* )} } \right) \\
 \end{array}
\end{equation}
From a value function point of view (21) we can be define the regret as:
\begin{equation}
{\rm{Regret}}(\pi ,\pi ^* ) = E\left[ {V^\pi  ({\textbf{\emph{S}}}) - V^{\pi ^* } ({\textbf{\emph{S}}})} \right]
\end{equation}
Where the expectation $E[\cdot]$ is with respect to the state $\textbf{\emph{S}}$. We give a regret bound of our learning algorithm following the assumption of the estimation errors $Q^{\pi}(\textbf{\emph{S}},\textbf{\emph{A}})-V^{*}(\textbf{\emph{S}})$ are uniformly random in $[-1,1]$ \cite{Thrun1993Issues}.
\begin{theorem}[Regret bound]
Consider a state $\textbf{\emph{S}}$ in which all the true optimal action values are equal at $Q^{*}(\textbf{\emph{S}},\textbf{\emph{A}})=V^{*}(\textbf{\emph{S}})$. Suppose that the estimation errors $Q^{\pi}(\textbf{\emph{S}},\textbf{\emph{A}})-V^{*}(\textbf{\emph{S}})$ are independently uniformly randomly in $[-1,1]$. Then,
\begin{equation}
{\rm{Regret}}(\pi ,\pi ^* )  \le 3
\end{equation}
\end{theorem}

\emph{Proof :} Please refer to Appendix C.

\section{Performance Evaluation}

In this section, simulation results are provided to illustrate the effectiveness of the proposed algorithm. There are another four solutions used for comparison. Specifically, actor-critic solution \cite{Bertsekas2011Neuro}, normalized advantage function (NAF) without deep neural network \cite{7989385}, normalized advantage function with deep neural network \cite{gu2016continuous} and random rate allocation solution. We show the performances in terms of the buffer size and the maximum network capacity. The Zipf distribution is used in our scenario, which is used to describe the popularity of the contents in user requirements. This distribute predicts that out of a population of $N$ elements, the frequency of elements of rank $k$ is:
\begin{equation}
f(k,z,N) = \frac{{1/k^z }}{{\sum\limits_{i = 1}^N {(1/i^z )} }}
\end{equation}
where $z$ is the zipf factor, in our scenario $z = 0.8$ \cite{4801529}. The default parameter values are shown in Table I.
\begin{table}
\captionsetup{format=plain, labelfont=bf,
  justification=raggedright, labelsep=newline}
\caption{SIMULATION PARAMETERS}

  \centering
  \begin{tabular}{lc}
  \hline
  Parameter & Value \\
  \hline
  Number of base station $F$ & 20\\
  The length of each video chunk $\Delta T_m$ (s)  & 10 \\
  Available rates (Mb/s)  & $\{ 0\}  \cup [2,10]$ \\
  Capacity space (Mb/s) & $[2,80]$ \\
  Buffer size $U$ (Mb) & 180 \\
  Random process \({{\cal N}}\) & Gauss process \\
  Average sojourn time $\mathbb{E} [\Omega _{j}]$ (s) & 60  \\
  \hline
  \end{tabular}
\end{table}

Fig. 4 shows the training process of our deep self-transfer reinforcement learning. The training clock is 1500 means that the self-transfer neural network run 1500 times. Then the deep reinforcement neural network start work. We can find that compared with the deep neural network without STL, using knowledge transfer make our agent get higher rewards.

We gain insight into the convergence behaviors of our approach. We compare it to other methods in terms of performance. Fig. 2 shows the convergence behavior of our algorithm. After 2000 training, the agent has mastered the knowledge of the whole network system. We can find that our proposed algorithm is obviously better than the another four methods.
\begin{figure}
\centering
\resizebox{6.8cm}{5.0cm}{\includegraphics{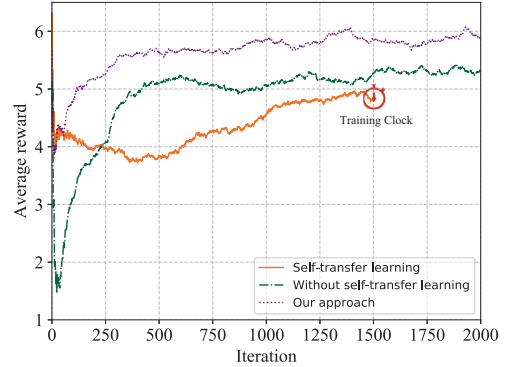}}
\caption{Training of deep self-transfer reinforcement learning.}
\end{figure}

\begin{figure*}
  \begin{minipage}[t]{0.5\linewidth}
    \centering
    \includegraphics[width=2.6in,height=1.8in]{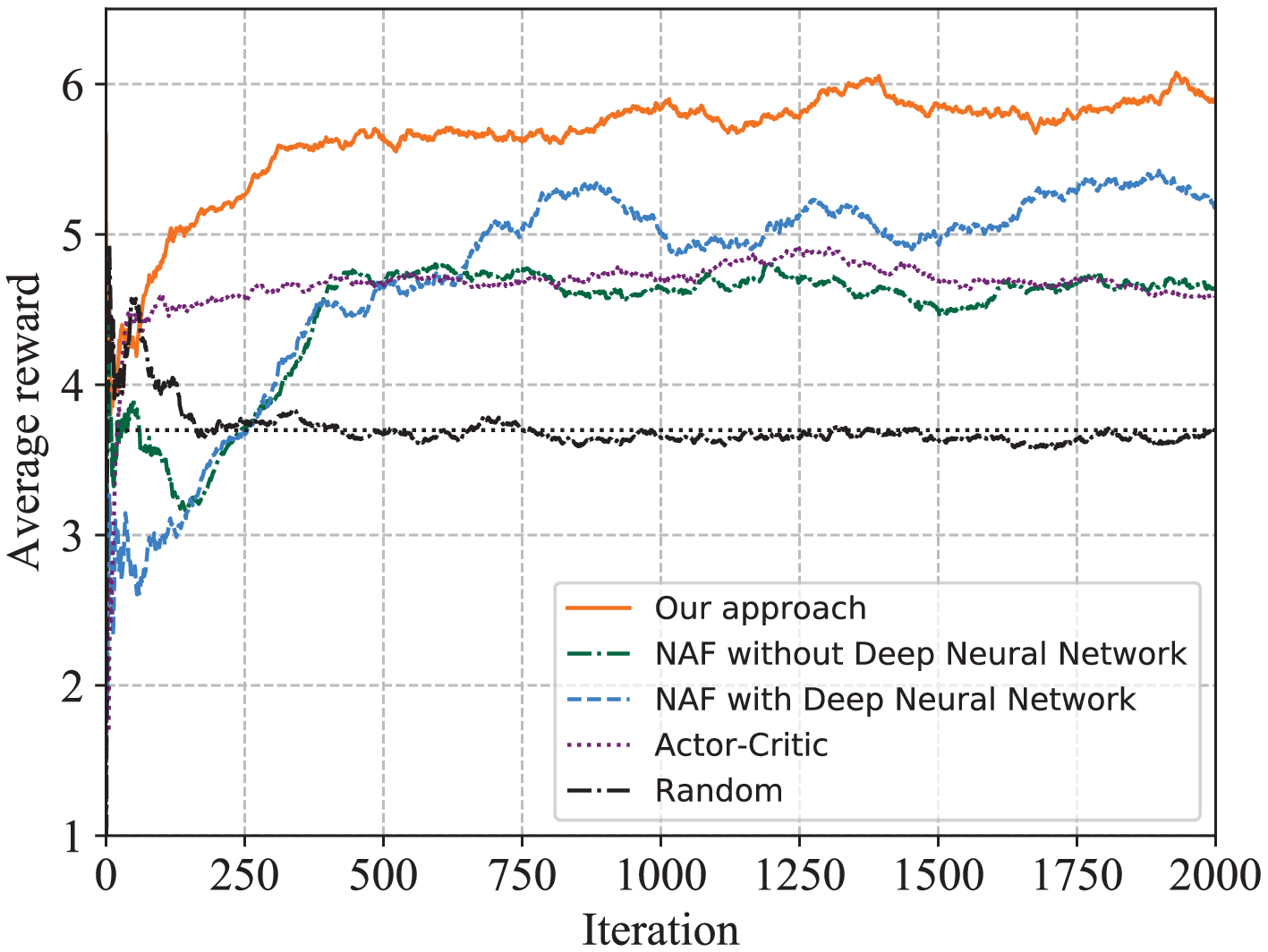}
    \caption{Convergence behavior.}
  \end{minipage}%
  \begin{minipage}[t]{0.5\linewidth}
    \centering
    \includegraphics[width=2.6in,height=1.8in]{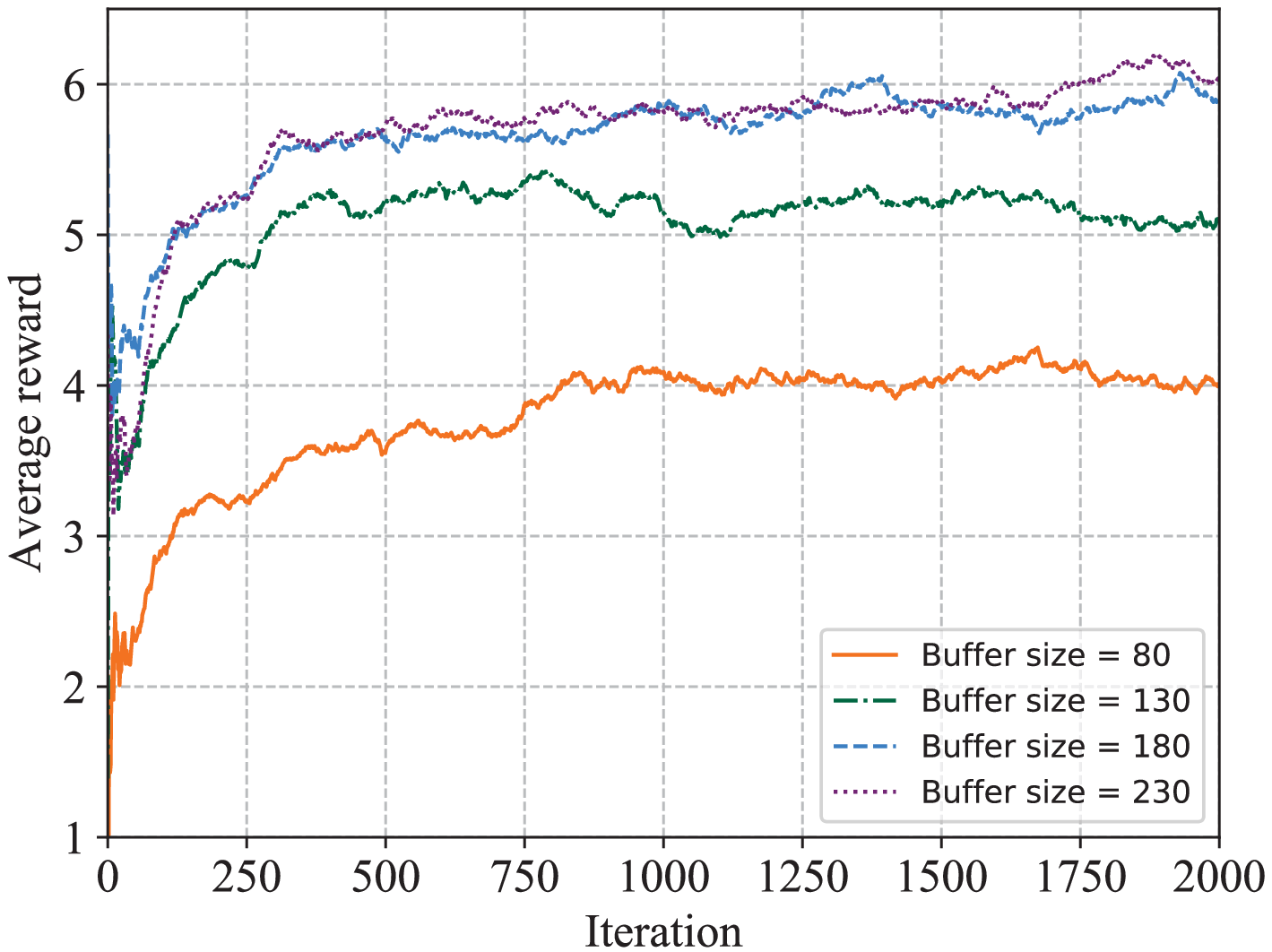}
    \caption{The impact of buffer size on our algorithm}
  \end{minipage}%
%  \begin{minipage}[t]{0.25\linewidth}
%    \centering
%    \includegraphics[width=2in,height=1.6in]{Fig4.eps}
%    \caption{Performance comparison with different buffer size}
%  \end{minipage}%
%  \begin{minipage}[t]{0.25\linewidth}
%    \centering
%    \includegraphics[width=2in,height=1.6in]{Fig5.eps}
%    \caption{Performance comparison with different maximum network capacity}
%  \end{minipage}%
\end{figure*}

\begin{figure*}
  \begin{minipage}[t]{0.5\linewidth}
    \centering
    \includegraphics[width=2.6in,height=1.8in]{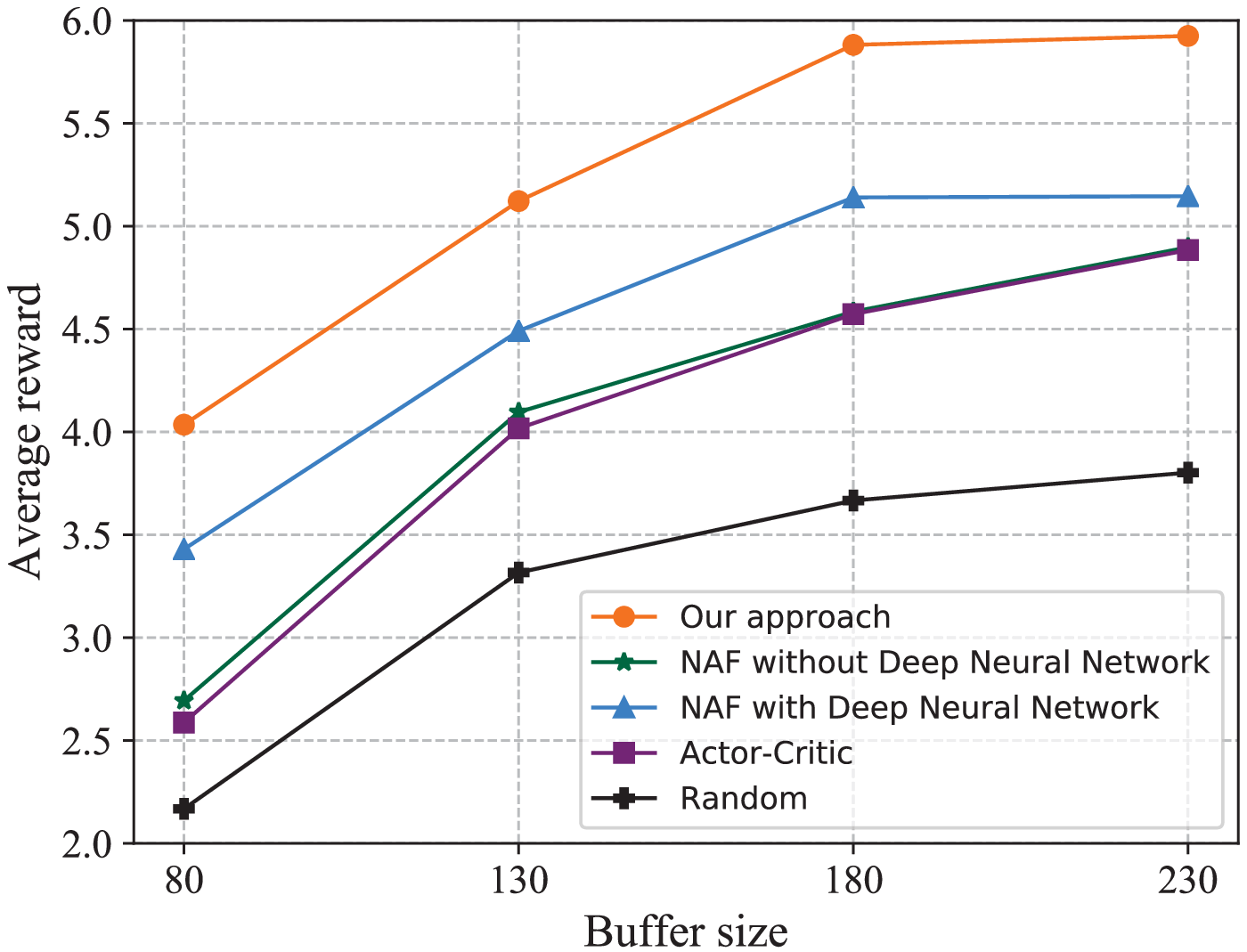}
    \caption{Performance comparison with different buffer size}
  \end{minipage}
  \begin{minipage}[t]{0.5\linewidth}
    \centering
    \includegraphics[width=2.6in,height=1.8in]{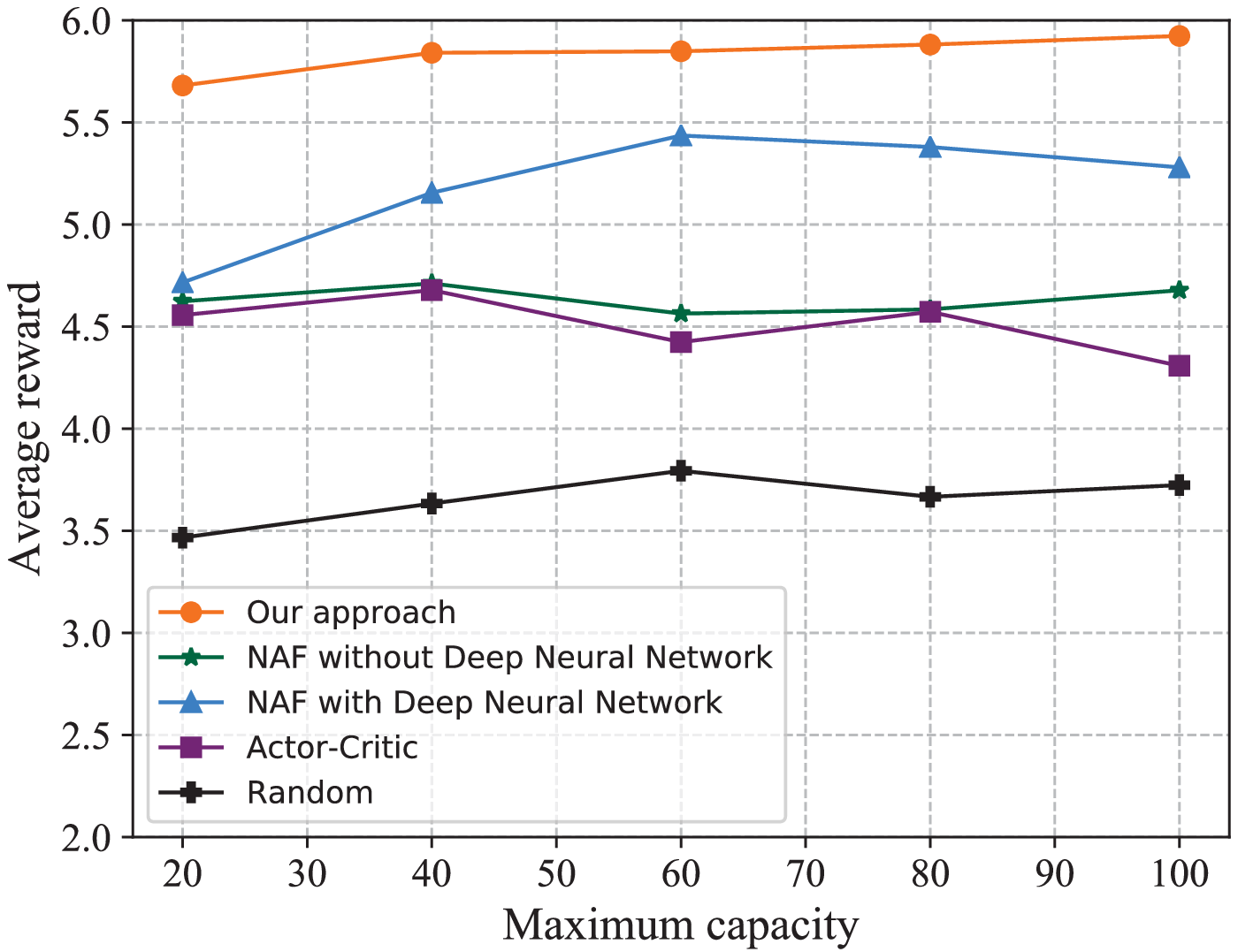}
    \caption{Performance comparison with different maximum network capacity}
  \end{minipage}
\end{figure*}

We let the buffer size change from 80Mb to 230Mb to compare the effect of the buffer size of the five methods. Fig. 3 demonstrates the impact of the buffer size on our approach. It can be found that as the buffer size increases, the average reward gained by the user is increasing because buffer size becomes larger means that more video chunks can be downloaded at the appropriate time. However, it is noticed that this performance improvement is becoming smaller as the buffer size becomes larger, which means that the average reward tends to converge as the buffer size increases. We can see from Fig. 4 that in the process of increasing buffer size, our algorithm is always better than the other four algorithms.

Fig. 5 demonstrates the impact of the maximum network capacity on the five approaches. In order to compare the performance of different algorithms, we set the user's storage space as 180Mb. It is shown that as the maximum network capacity increase, our algorithm has always been able to maintain a higher QoE than the another three methods it can also maintain a higher QoE when the maximum network capacity is low.

\section{Conclusion}
Existing researches on cache-enabled video rate adaptive allocation do not take into account the continuous system capacity and candidate rate. In this paper, we leverage MDP to model the cache-enabled video rate allocation problem and propose a QoE-driven  dynamic video rate adaptation method via deep continuous Q-learning scheme. We study the impacts of the buffer size and the maximum network capacity on our approach, and the performance evaluation results have shown that our approach can greatly improve the QoE. Future work is going to consider a more efficient scheme by adding more prior information to the model.

\appendices
\section{Proof of Lemma 1}
By Propositions 4.2.1, 4.2.3, and 4.2.5 in \cite{Bertsekas:1995:DPO:526593}, the optimal system reward of problem (10) is the same for all initial states and the solution $(\delta, V\left( \textbf{\emph{S}}^{t_x } \right))$ to the                                                                                            following Bellman equation exists.
\begin{equation}
\begin{split}
& \delta  + V\left( \textbf{\emph{S}} \right)=
\\&\mathop {\max }\limits_{\textbf{\emph{A}} \in \cal A } \left\{ {\pi(\textbf{\emph{S}}|\textbf{\emph{A}})(r^{t_x}_{\rm QoE} +\gamma^{t_x}\sum_{\textbf{\emph{S}}^{' } \in \cal {S} }p(\textbf{\emph{S}}^{' }|\textbf{\emph{S}},\textbf{\emph{A}}^)V(\textbf{\emph{S}}^{' }))} \right\}
\end{split}
\end{equation}
The transition probability is given by
\begin{equation}
\begin{split}
&p(\textbf{\emph{S}}^{' }|\textbf{\emph{S}},\textbf{\emph{A}}^)\buildrel \Delta \over =p(\textbf{\emph{S}}^{t_{x+1} }=\textbf{\emph{S}}^{' }|\textbf{\emph{S}}^{t_x }=\textbf{\emph{S}},\textbf{\emph{A}}^{t_x }=\textbf{\emph{A}})\\
&=E_{n,C}[p(\textbf{\emph{S}}^{t_{x+1} }=\textbf{\emph{S}}^{' }|\textbf{\emph{S}}^{t_x }=\textbf{\emph{S}},\textbf{\emph{A}}^{t_x }=\textbf{\emph{A}}),n^{t_x}=n,C^{t_x}=C]
\end{split}
\end{equation}
Substituting (31) into (30) leads to (11).
\section{Proof of Theorem 1}
Since $n^{t_x}$ and $C^{t_x}$ affect the state transition probabilities, we let $d^{\mu}(\textbf{\emph{S}})=\sum_{x=0}^{\infty}\gamma^{t_x}p(\textbf{\emph{S}}^{t_x}=\textbf{\emph{S}}|\textbf{\emph{S}}^{t_0},\textbf{\emph{A}})$. Then use policy gradient \cite{Sutton1999Policy} and cost function $L$ (line 14 in Algorithm 3) we get
\begin{equation}
\frac{{\partial L}}{{\partial \theta }} = \sum\limits_{\textbf{\emph{S}}} {d^\mu  (\textbf{\emph{S}})} \sum\limits_{\textbf{\emph{A}}} {\frac{{\partial \mu (\textbf{\emph{S}},\textbf{\emph{A}})}}{{\partial \theta }}[A^{\mu}(\textbf{\emph{S}},\textbf{\emph{A}}) + V^{\mu}(\textbf{\emph{S}})]}
\end{equation}
The gradient is no terms of the form $\frac{{\partial d^\mu  (\textbf{\emph{S}})}}{{\partial \theta }}$, so the effect of policy changes on the distribution of states does not appear.
\section{Proof of Theorem 2}
Define $\sigma _\textbf{\emph{A}}  = Q^\pi  (\textbf{\emph{S}},\textbf{\emph{A}}) - Q^{\pi ^* } (\textbf{\emph{S}},\textbf{\emph{A}})$. $\sigma _{\textbf{\emph{A}}}$ is a uniform random variable in $[-1,1]$. As we us the Sigmoid function as the activation function, the output of $\mu(\textbf{\emph{S}})$ belong to $[0,1]$. We quantize the interval $[0,1]$ into $m$ values at the interval $1/m$. Thus we get $m$ discrete actions. When $m$ tends to infinity we can get continuous actions. Because the estimation errors are independent, we can derive
\begin{equation}
\begin{split}
 P(\mathop {\max }\limits_{\textbf{\emph{A}}} \sigma _{\textbf{\emph{A}}}  \le \sigma ) & = P(\sigma _1  \le \sigma ,\sigma _2  \le \sigma ,...,\sigma _m  \le \sigma ) \\
  & = \prod\limits_{i = 1}^m {P(\sigma _i  \le \sigma )}  \\
\end{split}
\end{equation}
The function $P(\sigma _i  \le \sigma )$ is the cumulative distribution function of $\sigma _i$, which here is defined as
\begin{equation}
P(\sigma _i  \le \sigma ) = \left\{ \begin{array}{l}
 0,{{if }} \; \sigma  \le 1 \\
 (1 + \sigma )/2,{{if }} \; 0{\rm{ < }}\sigma  < 1 \\
 1,{{if }} \; \sigma  \ge 1 \\
 \end{array} \right.
\end{equation}
Then we get that
\begin{equation}
\begin{array}{l}
\begin{split}
 P(\mathop {\max }\limits_{\textbf{\emph{A}}} \sigma _{\textbf{\emph{A}}}  \le \sigma ) & = \prod\limits_{i = 1}^m {P(\sigma _i  \le \sigma )}  \\
  & = \left\{ \begin{array}{l}
 0,if\sigma \; \le  - 1 \\
 \left[ {(1 + \sigma )/2} \right]^m ,if \; - 1 < \sigma  < 1 \\
 1,if \; \sigma  > 1 \\
 \end{array} \right. \\
 \end{split}
 \end{array}
\end{equation}
The expectation of $\mathop {\max }\limits_{\textbf{\emph{A}}} \sigma _{\textbf{\emph{A}}}$ can be given as
\begin{equation}
E[\mathop {\max }\limits_{\textbf{\emph{A}}} \sigma _{\textbf{\emph{A}}} ] = \int\limits_{ - 1}^1 {\sigma f_\sigma  (\sigma )d\sigma }
\end{equation}
where $f_\sigma$ is the probability density function of $\sigma$, it can be written as $f_\sigma  (\sigma ) = \frac{m}{2}\left[ {(1 + \sigma )/2} \right]^{m - 1}$. Note that we use Sigmoid function as the activation function of the $\mu \left( {\textbf{\emph{S}}}\right)$ and $\textbf{\emph{L}}(\textbf{\emph{S}})$. Thus
\begin{equation}
\textbf{\emph{P}}(\textbf{\emph{S}})=\textbf{\emph{L}}(\textbf{\emph{S}})^{T} \textbf{\emph{L}}(\textbf{\emph{S}})\leq 1.
\end{equation}
\begin{equation}
A^{\pi}\left( {{\textbf{\emph{S}}},{\textbf{\emph{A}}} } \right) =  - \frac{1}{2}\left( {{\textbf{\emph{A}}} - \mu \left( {{\textbf{\emph{S}}} } \right)} \right)^T {\textbf{\emph{P}}}\left( {{\textbf{\emph{S}}} } \right)\left( {{\textbf{\emph{A}}} - \mu \left( {{\textbf{\emph{S}}} } \right)} \right) \geq -2
\end{equation}
This implies that
\begin{equation}
\begin{array}{l}
 E\left[ {V^\pi  (\textbf{\emph{S}}) - V^{\pi ^* } (\textbf{\emph{S}})} \right] \\
  = E\left[ {Q^\pi  (\textbf{\emph{S}},\textbf{\emph{A}}) - A^\pi  (\textbf{\emph{S}}) - V^{\pi ^* } (\textbf{\emph{S}})} \right] \\
  \le E\left[ {Q^\pi  (\textbf{\emph{S}},\textbf{\emph{A}}) - V^{\pi ^* } (\textbf{\emph{S}})} \right] + 2 \\
  = E[\mathop {\max }\limits_\textbf{\emph{A}} \sigma _\textbf{\emph{A}} ] + 2 \\
  = \left[ {\left[ {(1 + \sigma )/2} \right]^m \frac{{m\sigma  - 1}}{{m + 1}}} \right]\left| \begin{array}{l}
 ^1  \\
 _{ - 1}  \\
 \end{array} \right. + 2 \\
  = \frac{{m - 1}}{{m + 1}} + 2 \\
 \end{array}
\end{equation}
In our cache-enabled bit rate allocation problem, the actions lie in the continuous space thus when $m \to \infty$
\begin{equation}
E\left[ {V^\pi  (S) - V^{\pi ^* } (S)} \right] \le 3
\end{equation}
We complete the proof.


%\section*{Acknowledgment}



\bibliographystyle{IEEEtran}
\bibliography{IEEEabrv,BiteRate}
%\begin{thebibliography}{1}
%
%
%%\bibitem{IEEEhowto:kopka}
%%H.~Kopka and P.~W. Daly, \emph{A Guide to \LaTeX}, 3rd~ed.\hskip 1em plus
%%  0.5em minus 0.4em\relax Harlow, England: Addison-Wesley, 1999.
%
%\end{thebibliography}

\end{document}